\documentclass[12pt]{article}
\usepackage{amscd}
\usepackage{verbatim}
\usepackage{amssymb}

\begin{document}
\vskip 0 true cm \flushbottom
\begin{center}
\vspace{24pt} { \large \bf A Metric for Gradient RG Flow of the
Worldsheet Sigma Model Beyond First Order
} \\

\vspace{30pt}
{\bf T Oliynyk}$^{\dag}$ \footnote{todd.oliynyk@aie.mpg.de}
\footnote
{Address after 1 June 2007: School of Mathematical Sciences, Monash
University, Vic 3800, Australia.}
,
{\bf V Suneeta}$^{\sharp}$ \footnote{suneeta@math.unb.ca} \footnote
{Address after 1 July 2007: Dept of Mathematical and Statistical
Sciences, University of Alberta, Edmonton AB, Canada T6G 2G1.}
{\bf E Woolgar}$^{\flat}$ \footnote{ewoolgar@math.ualberta.ca} 

\vspace{24pt} 
{\footnotesize $^\dag$ Max-Planck-Institut f\"ur Gravitationsphysik
(Albert Einstein Institute), Am M\"uhlenberg 1,
D-14476 Potsdam, Germany.\\
$^\sharp$ Dept of Mathematics and Statistics, University of New
Brunswick, Fredericton, NB, Canada E3B 5A3.\\
$^\flat$ Dept of Mathematical and Statistical Sciences, University of Alberta,\\
Edmonton, AB, Canada T6G 2G1.}
\end{center}
\date{\today}
\bigskip

\begin{center}
{\bf Abstract}
\end{center}
Tseytlin has recently proposed that an action functional exists
whose gradient generates to all orders in perturbation theory the
Renormalization Group (RG) flow of the target space metric in the
worldsheet sigma model. The gradient is defined with respect to a
metric on the space of coupling constants which is explicitly known
only to leading order in perturbation theory, but at that order is
positive semi-definite, as follows from Perelman's work on the Ricci
flow. This gives rise to a monotonicity formula for the flow which
is expected to fail only if the beta function perturbation series
fails to converge, which can happen if curvatures or their
derivatives grow large. We test the validity of the monotonicity
formula at next-to-leading order in perturbation theory by
explicitly computing the second-order terms in the metric on the
space of coupling constants. At this order, this metric is found not
to be positive semi-definite. In situations where this might spoil
monotonicity, derivatives of curvature become large enough for
higher order perturbative corrections to be significant.

\newpage

\section{Introduction}
\setcounter{equation}{0}

\noindent It has been known for quite a long time that the
renormalization group (RG) flow of 2-dimensional nonlinear sigma
models, computed to first order in the loop expansion and neglecting
all but gravity, is a gradient flow generated by the
Einstein-Hilbert action. This first order RG flow \cite{Friedan} is
the Ricci flow, which can be written as
\begin{equation}
\frac{\partial g_{ij}}{\partial t} = -\alpha'R_{ij} ={\rm Grad\ }
\left [ \alpha'\int_M R dV\right ] \ . \label{eq1.1}
\end{equation}
Here we take $g_{ij}$ to be a Riemannian metric, $dV$ to be the
metric volume element, $R=g^{ij}R_{ij}$ to be the scalar curvature
of the metric, $t$ to be the logarithm of the renormalization scale,
and $\alpha'>0$ to be a constant,\footnote
{This constant is usually set equal to $2$ in the mathematics
literature.}
the string length squared, which serves as an expansion parameter in
the sigma model context.

The gradient here is on the ``space of coupling constants'', which
we take to be the space whose points represent positive symmetric
2-tensor fields on a manifold $M$. The inner product of the gradient
vector with another vector is a directional derivative which, in the
present context, is the first variational derivative of an ``action
functional'' or potential that generates the gradient flow (for
greater detail, see Section IV).

Now the variational derivative of the Einstein-Hilbert action
\begin{equation}
S_{\rm EH}:=\int_M RdV \label{1.2}
\end{equation}
on a closed manifold $M$ (so there are no boundary terms)\footnote
{Throughout we take $(M,g)$ to be a closed Riemannian manifold.}
in the direction $\frac{\partial g_{ij}}{\partial s}$ gives the very
familiar result:
\begin{eqnarray}
\frac{dS_{\rm EH}}{ds} &=& -\int_M
\left (R^{ij}-\frac{1}{2} g^{ij}R \right )
\frac{\partial g_{ij}}{\partial s}dV\nonumber\\
&=&-\int_M R_{ij}\left ( g^{ik}g^{jl}-\frac{1}{2}g^{ij}g^{kl} \right
) \frac{\partial g_{kl}}{\partial s}dV\ . \label{eq1.3}
\end{eqnarray}
If the metric were $<u,v>=\int_M u_{ij} v_{kl} g^{ik} g^{jl} dV$
then the gradient would be the negative of the Einstein tensor, but
if the metric is
\begin{equation}
\langle u , v \rangle := \int_M u_{ij}\left (
g^{ik}g^{jl}-\frac{1}{2}g^{ij}g^{kl} \right ) v_{kl}dV \ ,
\label{eq1.4}
\end{equation}
then the gradient is indeed the negative of the Ricci tensor
\cite{Gibbons}, verifying the second equality in (\ref{eq1.1}), and
giving the formula
\begin{equation}
\frac{dS_{\rm EH}}{dt} = \alpha'^2 \langle {\rm Ric}, {\rm Ric}
\rangle \label{1.5}
\end{equation}
for the derivative of the action along the flow. If the metric
$\langle \cdot , \cdot \rangle$ were positive semi-definite, this
formula would show that the action would increase monotonically
along the flow, but obviously this metric is not of definite sign.
As a result, the gradient can in principle change between being
``timelike'' and being ``spacelike'' according to whether the trace
or tracefree part of the Ricci tensor dominates. Along any flow for
which such a change occurs, the Einstein-Hilbert action will not be
a monotonic function of the flow parameter.

The apparent lack of a monotonicity formula along the RG flow is
surprising in view of the Zamolodchikov $C$-theorem \cite{Z}, which
guarantees a monotonic quantity along RG flow for a 2-dimensional
unitary quantum field theory with a {\em finite} number of couplings
(as opposed to the current case, where the coupling constants,
usually found by expanding $g_{ij}(x)$ around a point  $x_0 \in M$ are
infinite in number). For a
discussion of the problems associated with generalizing the
$C$-theorem to the worldsheet sigma model (on a curved worldsheet),
we refer the reader to the summary in \cite{Tseytlin}.

There is, however, another approach which does yield a monotonicity
formula for first order RG flow and possibly beyond. In his
celebrated work on Ricci flow, Perelman \cite{Perelman} has proposed
an approach based on enlarging the space of coupling constants to
include an extra function which then generates diffeomorphisms that
act by pullback on $g_{ij}$. A choice of this function gives a
submanifold of the enlarged space onto which the original space of
coupling constants can be mapped, and can be thought of as a choice
of parametrization of the coupling constants $g_{ij}$ in the sigma
model. The first order RG flow induces a flow on this submanifold,
and the submanifold can be chosen so that the induced flow is
gradient with respect to a {\it positive definite} metric. The
submanifold is selected in a very natural way: one fixes the extra
function above to be given by the lowest eigenfunction of a certain
Schr\"odinger problem\footnote
{A special case of this Schr\"odinger problem first appeared in the
study of RG flows in \cite{FOZ}, which studied the case of a
2-dimensional target space.}
on the manifold $(M,g_{ij})$. We have described this construction in
greater detail in \cite{OSW}.

While Perelman's approach works to first order in $\alpha'$, there
remains the question of whether the full RG flow is gradient with
respect to a positive definite metric. Tseytlin has recently
addressed this question \cite{Tseytlin}. He starts with an action
functional which is the integral over the target space of the
``generalized central charge function'', a particular combination of
metric and dilaton $\beta$-functions discussed in \cite{CKP,
Tseytlin1}, to which he appends a Lagrange multiplier term. Upon
truncating the generalized central charge to first order in
$\alpha'$ and extremizing the resulting action functional with
respect to the dilaton, one can reproduce Perelman's construction,
so the first order RG flow of the target space metric is obtained as
a gradient flow of the truncated action functional of Tseytlin. Then
Tseytlin invokes results of Osborn \cite{Osborn} to argue that the
{\em untruncated} gradient generates {\em to all orders in
perturbation theory} the RG flow of the sigma model's target space
metric. \footnote
{In the process, the dilaton becomes metric dependent (it in fact
satisfies the equation of the lowest eigenfunction of a
Schr\"odinger operator describing the wavefunction of a particle
coupled to gravity via the curvature scalar). This dilaton no longer
satisfies its own independent RG flow equation.}

The corresponding metric on the space of coupling constants is not
explicitly given beyond first order in \cite{Tseytlin} (to that
order it is just the metric obtained from Perelman's construction
\cite{Perelman, OSW}). Thus the issue of monotonicity of this action
functional under RG flow beyond first order remains to be explored.
Tseytlin argues that a strict monotonicity formula is not necessary.
Rather, since the leading (Perelman) term in the derivative of the
action along the flow is positive, failure of monotonicity indicates
that higher order terms become dominant. This suggests that perhaps
the perturbation series for the $\beta$-functions will fail to
converge whenever monotonicity of the action fails; conversely,
monotonicity holds whenever perturbation theory makes sense. A motivation
for this expectation is the fact that the central charge action is
related to the Zamolodchikov $C$-function, and upon applying
Perelman's construction, the hope is that it indeed {\em behaves} like
a $C$-function, and is monotonic under RG flow to all orders.

It is difficult to test this since the full perturbation series is
not known explicitly. However, we take a pragmatic view. Say the
$\beta$-functions are known to some order $p$. Then the central
charge action (plus Lagrange multiplier term) is also known at this
order, and one can compute its derivative along the flow and check
for monotonicity. This will reveal the circumstances $C$, if any, in
which monotonicity may fail at order $p$. If $C$ is non-empty, one
can then attempt to estimate whether the order $p$ truncation of the
$\beta$-functions is valid or whether higher-order, neglected terms
are, in circumstances $C$, comparable in size to the lower-order,
untruncated terms. If so, {\em the order $p$ truncation breaks
down}; i.e., the truncation should be extended. The view in
\cite{Tseytlin} would be confirmed if such an extension either
restores monotonicity or eventually points to a divergent
perturbation series, but these are not the only possible outcomes. A
reliable assessment would require greater knowledge of the
perturbation series than is presently available.

The purpose of the present work is to confirm that the issue does
arise, because the metric that emerges from the proposal in
\cite{Tseytlin} is not order-by-order of definite sign; indeed, the
issue will arise at second order in $\alpha'$.

There are essentially two ways in which truncations at finite order
and perturbation theory may become unreliable. Judging from the
known terms in the perturbation series for $\beta$ (e.g.,
\cite{JJM}), these are when either (i) curvatures become large
($\sim 1/\alpha'$ or larger), or (ii) derivatives of curvatures
become large. The problem can occur even when the curvature is small
in magnitude, if some derivative of curvature is sufficiently
large.\footnote
{One may suggest that RG flow will smooth out the inhomogeneities
that generate large derivatives. This is not always clear. Ricci
flow, for example, does not always smooth out inhomogeneities.}

Let us now look more closely at the mechanism by which monotonicity
might fail when passing from leading order in $\alpha'$ to next
order. If $S$ is the action and RG flow is its gradient flow, then
schematically at least, along the flow we have
\begin{eqnarray}
\frac{dS}{dt}&=&\kappa(\beta,\beta)\ , \label{eq1.6}\\
&=&\int_M \biggl [ \kappa^{ijkl}_{(0)} \left ( \beta^{(1)}_{ij}
\beta^{(1)}_{kl}+\beta^{(1)}_{ij}\beta^{(2)}_{kl}+\beta^{(2)}_{ij}
\beta^{(1)}_{kl}+\dots\right )\nonumber\\
&&\qquad+\kappa^{ijkl}_{(1)} \beta^{(1)}_{ij} \beta^{(1)}_{kl}
+\kappa^{ijklmn}_{(1)}
\nabla_m\beta^{(1)}_{ij}\nabla_n\beta^{(1)}_{kl} +\dots\biggr ]dm\ .
\label{eq1.7}
\end{eqnarray}
Here $\kappa(\cdot,\cdot)$ is the metric on the space of coupling
constants, $dm$ is some measure, and $\beta$ represents the
$\beta$-function for the target space metric. The subscript or
superscript in parentheses indicates the order in $\alpha'$, so we
keep only terms up to order $\alpha'^3$ inclusive (the leading term
being of order $\alpha'^2$). On dimensional grounds, higher
derivatives than those shown cannot occur at this order. Since
truncation at leading order is just the case studied in
\cite{Perelman}, we see that $\kappa^{ijkl}_{(0)}$ is positive
semi-definite. Monotonicity at next-to-leading order becomes a
question of the signatures of the two $\kappa_{(1)}$ coefficients.

We will confirm by explicit variation of the second-order action
that to second order in $\alpha'$ the RG flow is the gradient flow
of Tseytlin's action functional and that its flow derivative has the
form (\ref{eq1.7}) (with the diffeomorphism-improved
$\beta$-function ${\bar \beta}_{ij}$, defined in (\ref{eq1.8}),
appearing in place of $\beta$ above). Furthermore,
$\kappa^{ijkl}_{(1)}=0$, but $\kappa^{ijklmn}_{(1)}$ is of
indefinite sign so $\kappa$ (truncated at order $\alpha'$) is no
longer positive semi-definite and so the RG flow, truncated at
second order, does not have a monotonicity formula. This happens
precisely in situation (ii) above; i.e., when first derivatives of
the curvatures are as large as ${\cal O}(|{\rm
Riem}|/\sqrt{\alpha'})$, and may signal a breakdown in perturbation
theory. Interestingly, large and even arbitrarily large curvatures
will not violate monotonicity at second order if the curvature is
sufficiently homogeneous---even though for large enough curvatures
the sigma model perturbation theory certainly breaks down.

We find in particular that on Ricci solitons the monotonicity
formula holds for the second order RG flow. Indeed, monotonicity
holds at second order on a wider class of metrics than solitons,
namely those with harmonic curvature operator. This condition is not
preserved along the second-order flow, so monotonicity along a flow
that begins at a metric with harmonic curvature can eventually break
down at large enough $t$ along the flow.

We follow \cite{Tseytlin} for our definitions of $\beta$-functions.
In particular, we choose local coordinates on $M$ so that the RG
flow of the target space metric $g_{ij}$ and dilaton $\phi$ is
expressed as
\begin{eqnarray}
\frac{\partial g_{ij}}{\partial t}&=& -{\bar \beta}^g_{ij} =
-\alpha' \left ( R_{ij} +2\nabla_i\nabla_j\phi\right ) -
\frac{\alpha'^2}{2}R_{iklm}R_j{}^{klm}+{\cal O}(\alpha'^3) \ ,
\label{eq1.8}\\
\frac{\partial \phi}{\partial t}&=&-{\bar \beta}^{\phi}
=-c_0+\alpha'\left (\frac{1}{2}\Delta \phi -\vert \nabla \phi
\vert^2 \right ) -\frac{\alpha'^2}{16} \vert {\rm Riem} \vert^2
+{\cal O}(\alpha'^3)\ . \label{eq1.9}
\end{eqnarray}

This paper is organized as follows. Section 2 reviews Tseytlin's
proposal and Perelman's technique. Section 3 extends the analysis to
second order in $\alpha'$. Section 4 shows that the second order
flow is gradient and contains the formula for the derivative along
the flow of Tseytlin's action ${\cal S}$. Section 5 contains a brief
discussion of metrics for which monotonicity does not break down. We
reiterate that, throughout, all manifolds are closed Riemannian
manifolds.

\section{Tseytlin's Proposed Potential}
\setcounter{equation}{0}

\noindent In this section, we review Tseytlin's proposal and the
result of Perelman upon which it is based.

Consider the ``central charge action" \cite{CKP, Tseytlin1},
modified by a Lagrange multiplier term:
\begin{eqnarray}
S(g,\phi)&:=&\int_M {\tilde \beta}^{\phi} e^{-2\phi}dV + \lambda \left
( \int_M e^{-2\phi}dV-1 \right ) \ , \label{eq2.1}\\
{\tilde \beta}^{\phi}&:=&{\bar \beta}^{\phi} -\frac{1}{4}g^{ij}{\bar
\beta}^g_{ij}\nonumber\\
&=& c_0 -\alpha' \left ( \Delta \phi -\vert \nabla \phi \vert^2
+\frac{1}{4}R \right )-\frac{\alpha'^2}{16} \vert {\rm Riem} \vert^2
+{\cal O}(\alpha'^3)\ . \label{eq2.2}
\end{eqnarray}
Tseytlin's proposal is that the RG flow for $g_{ij}$ is the gradient
of the action\footnote
{The sign convention for the action is opposite that of Perelman, so
the desired monotonicity property will be a monotone {\em
decrease}.}
\begin{equation}
{\cal S}(g):={\hat S}(g,\varphi)\ , \label{eq2.3}
\end{equation}
where
\begin{equation}
\varphi=-\log \Phi\label{eq2.4}
\end{equation}
and $\Phi$ solves the eigenvalue problem
\begin{eqnarray}
&&\alpha' \left ( \Delta - \frac{1}{4}R-\frac{\alpha'}{16} \vert
{\rm Riem} \vert^2+{\cal O}(\alpha'^2)\right )\Phi=
-(\lambda+c_0)\Phi\ , \label{eq2.5}
\\
&&1=\int_M \Phi^2 dV\equiv \int_M e^{-2\varphi}dV \ . \label{eq2.6}
\end{eqnarray}

In the action $\lambda$ appears as a Lagrange multiplier, and $c_0$
is a free parameter. Note that $c_0+\lambda$ must be the lowest
eigenvalue of the operator on the left-hand side of (\ref{eq2.5})
\footnote{ $\varphi$ is therefore sometimes called the {\em
minimizer}.}, since by (\ref{eq2.4}) $\Phi$ cannot have nodes;
otherwise the logarithm would fail to be defined. The eigenvalue
problem (\ref{eq2.4}--\ref{eq2.6}) arises by extremizing the action
${\hat S}(g, \phi)$ with respect to $\phi$ and $\lambda$. The
dilaton RG flow cannot be obtained as a gradient flow of
(\ref{eq2.3}) since the action ${\cal S}(g)$ is not a functional of
$\phi$.

It is easily checked that (\ref{eq2.2}--\ref{eq2.6}) imply
\begin{equation}
{\tilde \beta}^{\varphi}=-\lambda={\cal S}(g)\ . \label{eq2.7}
\end{equation}
where of course $\lambda$ depends nontrivially on $g$ due to
(\ref{eq2.5}).

An arbitrary one-parameter variation of the action (\ref{eq2.1})
yields
\begin{eqnarray}
\frac{dS}{ds}&=& \int_M \left [ -\frac{1}{4}{\bar \beta}_{ij}
\frac{\partial g^{ij}}{\partial s}-\frac{1}{4}g^{ij}\frac{\partial
{\bar \beta}_{ij}}{\partial s}-\frac{\partial {\bar \beta}^{\phi}}
{\partial s}\right ] e^{-2\phi}dV\nonumber\\
&&+\int_M \left ( {\tilde \beta}^{\phi}+\lambda \right )\frac{\partial}
{\partial s} \left ( e^{-2\phi} dV \right ) \nonumber \\
&&+\frac{\partial \lambda}{\partial s} \left ( \int_M e^{-2\phi}dV
-1 \right ) \label{eq2.8}
\end{eqnarray}
If we vary about the minimizer $\phi=\varphi$, then due to
(\ref{eq2.6}, \ref{eq2.7}) the last two integrals contribute
nothing. Thus (\ref{eq2.8}) reduces to
\begin{equation}
\frac{dS}{ds}= \int_M \left [ -\frac{1}{4}{\bar \beta}_{ij}
\frac{\partial g^{ij}}{\partial s}-\frac{1}{4}g^{ij}\frac{\partial
{\bar \beta}_{ij}}{\partial s}-\frac{\partial {\bar
\beta}^{\phi}}{\partial s}\right ]_{\phi=\varphi} e^{-2\varphi}dV \
. \label{eq2.9}
\end{equation}
Section 1 of \cite{Perelman} (see also \cite{OSW}) shows that if the
$\beta$-functions are replaced by their first-order truncations (at
the minimizer $\varphi$)
\begin{eqnarray}
{\bar \beta}^{(1)}_{ij}&=&\alpha' \left (
R_{ij}+2\nabla_i\nabla_j \varphi \right ) \ , \label{eq2.10} \\
{\bar \beta}^{\varphi(1)}&=& c_0-\alpha' \left ( \frac{1}{2}\Delta
\varphi -\vert \Delta \varphi \vert^2 \right ) \ , \label{eq2.11}
\end{eqnarray}
then the last two terms in the integrand vanish. One obtains simply
\begin{equation}
\frac{dS^{(1)}}{ds}= \frac{1}{4} \int_M g^{ik}g^{jl}{\bar
\beta}^{(1)}_{ij} \frac{\partial g_{kl}}{\partial s} \ ,
\label{eq2.12}
\end{equation}
so the first-order truncated flow
\begin{equation}
\frac{\partial g_{ij}}{\partial t} = -{\bar \beta}^{(1)}_{ij} \label{2.13}
\end{equation}
is clearly gradient, the metric is
\begin{equation}
(u,v)=\int_Mg^{ik}g^{jl} u_{ij}v_{kl} \ , \label{eq2.14}
\end{equation}
which is positive semi-definite, and along the flow we have the
monotonicity formula
\begin{equation}
\frac{dS^{(1)}}{dt} = -\frac{1}{4} \int_M \left \vert {\bar
\beta}^{(1)}_{ij} \right \vert^2 \ . \label{eq2.15}
\end{equation}
This implies that the derivative (\ref{eq2.9}) along the flow of the
full action has the form
\begin{equation}
\frac{dS}{dt}= -\frac{1}{4} \int_M \left ( \left \vert {\bar
\beta}_{ij} \right \vert^2 +{\cal O}(\alpha'^3) \right ) \ ,
\label{eq2.16}
\end{equation}
where the $\left \vert {\bar \beta}_{ij} \right
\vert^2$ term is ${\cal O} (\alpha'^2)$.

\section{Second Order Action}
\setcounter{equation}{0}

\noindent In this section, we include in the action the term
$\alpha'^2 |{\rm Riem}|^2$ which occurs in ${\tilde \beta}^{\phi}$,
and compute its variation. The result can also be reconstructed from
calculations in the literature (see \cite{JJM}). Readers wishing to
skip the routine calculational details may want to proceed straight
to the results (\ref{eq3.11}) and (\ref{eq3.12}).

For a one-parameter variation in the metric, where $s$ is the
parameter, we use the standard formul\ae
\begin{eqnarray}
\frac{\partial}{\partial s}R^i{}_{jkl}
&=&\nabla_k\frac{\partial}{\partial s}
\Gamma^i_{jl}-\nabla_l\frac{\partial}{\partial s}
\Gamma^i_{jk}\ , \label{eq3.1} \\
\frac{\partial}{\partial s} \Gamma^i_{jk} &=& \frac{1}{2}g^{il}\left
( \nabla_j \frac{\partial g_{lk}}{\partial s} + \nabla_k
\frac{\partial g_{jl}}{\partial s} - \nabla_l \frac{\partial
g_{jk}}{\partial s}\right )\ , \label{eq3.2}\\
\frac{\partial}{\partial s}dV&=&\frac{1}{2}g^{ij}\frac{\partial
g_{ij}}{\partial s}dV\ . \label{eq3.3}
\end{eqnarray}
Using these, we write
\begin{eqnarray}
&&\frac{\partial}{\partial s} \left [ -\frac{\alpha'^2}{16} \int_M
\vert {\rm Riem} \vert^2 e^{-2\phi} dV\right ] \nonumber \\
&=&-\frac{\alpha'^2}{16} \int_M \biggl [ 2R^p{}_{qrs} g_{pi} g^{qj}
g^{rk} g^{sl} \frac{\partial}{\partial s} R^i{}_{jkl}+
R^i{}_{klm}R^{jklm}\frac{\partial g_{ij}}{\partial s}\nonumber\\
&&\qquad + R^i{}_{klm} R_{ij}{}^{lm}\frac{\partial g^{jk}}{\partial
s}+ R^i{}_{klm}R_i{}^k{}_j{}^m \frac{\partial g^{jl}}{\partial s}
\nonumber\\
&&\qquad+R^i{}_{klm} R_i{}^{kl}{}_j \frac{\partial g^{mj}}{\partial
s}+\vert {\rm Riem} \vert^2\left ( \frac{1}{2} g^{ij} \frac{\partial
g_{ij}}{\partial s}-2\frac{\partial \phi}{\partial s}
\right ) \biggr ] e^{-2\phi} dV \nonumber\\
&=&-\frac{\alpha'^2}{16} \int_M \biggl [ 2R^{ijkl} \nabla_k \bigg (
\nabla_j \frac{\partial g_{il}}{\partial s} +\nabla_l \frac{\partial
g_{ij}}{\partial s}- \nabla_i \frac{\partial g_{jl}}{\partial
s}\bigg ) \nonumber\\
&&\qquad -2R^i{}_{klm}R^{jklm}\frac{\partial g_{ij}}{\partial s}+
\vert {\rm Riem} \vert^2\left( \frac{1}{2}g^{ij}\frac{\partial
g_{ij}}{\partial s}-2\frac{\partial
\phi}{\partial s} \right ) \biggr ] e^{-2\phi} dV\ .\nonumber\\
&& \label{eq3.4}
\end{eqnarray}
The term $R^{ijkl}\nabla_k\nabla_l\frac{\partial g_{ij}}{\partial
s}$ is easily seen by index symmetry to contribute zero, so we will
discard it. Next, we integrate by parts and use the second Bianchi
identity, once contracted, which shows that
\begin{equation}
\nabla_k R^{ijkl}=\nabla^i R^{jl} -\nabla^j R^{il}\ . \label{eq3.5}
\end{equation}
The result is
\begin{eqnarray}
&&\frac{\partial}{\partial s} \left [ -\frac{\alpha'^2}{16} \int_M
\vert {\rm Riem} \vert^2 e^{-2\phi} dV\right ] \nonumber\\
&=&-\frac{\alpha'^2}{16}\int_M \biggl [ 2\left ( \nabla^j R^{il} -
\nabla^i R^{jl} \right ) \left ( \nabla_j\frac{\partial
g_{il}}{\partial s}
-\nabla_i\frac{\partial g_{jl}}{\partial s}\right ) \nonumber\\
&&\quad + 4R^{ijkl}\nabla_k \phi \left ( \nabla_j\frac{\partial
g_{il}}{\partial s} - \nabla_i\frac{\partial g_{jl}}{\partial s}
\right ) -2R^i{}_{klm}R^{jklm}\frac{\partial g_{ij}}{\partial s}
\nonumber\\
&&\quad + \vert {\rm Riem} \vert^2 \left ( \frac{1}{2} g^{ij}
\frac{\partial g_{ij}}{\partial s} - 2\frac{\partial \phi}{\partial
s} \right ) \biggr ] e^{-2\phi} dV\ . \label{eq3.6}
\end{eqnarray}
We can  replace the $R^{ijkl}\nabla_k \phi$ term using the Ricci
identity
\begin{equation}
R^{ijkl}\nabla_k \phi= -\left ( \nabla^i\nabla^j-\nabla^j\nabla^i
\right ) \nabla^l \phi \ . \label{eq3.7}
\end{equation}
Finally, if we vary about the minimizer $\phi=\varphi$, then $\left
( \frac{1}{2} g^{ij} \frac{\partial g_{ij}}{\partial s} -
2\frac{\partial \phi}{\partial s} \right )$ vanishes. Using these
results, we obtain
\begin{eqnarray}
&&\frac{\partial}{\partial s} \left [ -\frac{\alpha'^2}{16} \int_M
\vert {\rm Riem} \vert^2 e^{-2\phi} dV\right ] \nonumber\\
&=&-\frac{\alpha'^2}{16}\int_M \biggl \{ 2\left [ \nabla^j \left (
R^{il} +2\nabla^i\nabla^l \phi \right ) - \nabla^i \left ( R^{jl}
+2\nabla^j\nabla^l \phi \right ) \right ]  \nonumber\\
&&\qquad \left ( \nabla_j\frac{\partial g_{il}}{\partial s}
-\nabla_i\frac{\partial g_{jl}}{\partial s}\right )
-2R^i{}_{klm}R^{jklm}\frac{\partial g_{ij}}{\partial s} \biggr \}
e^{-2\phi} dV\nonumber\\
&=&\frac{\alpha'}{8}\int_M \left ( \nabla^j {\bar \beta}^{(1)il} -
\nabla^i {\bar \beta}^{(1)jl} \right ) \left (
\nabla_j\frac{\partial g_{il}}{\partial s} -\nabla_i\frac{\partial
g_{jl}}{\partial
s}\right )\bigg \vert_{\phi=\varphi} e^{2\varphi}dV\nonumber\\
&&+\frac{1}{4}\int_M {\bar\beta}^{(2)ij}\frac{\partial
g_{ij}}{\partial s} e^{-2\varphi} dV \ , \label{eq3.8}
\end{eqnarray}
where we define
\begin{equation}
{\bar \beta}^{(2)}_{ij}=\frac{\alpha'}{2} R_{iklm}R_j{}^{klm} \label{eq3.9}
\end{equation}
so that
\begin{equation}
{\bar \beta}_{ij}={\bar \beta}^{(1)}_{ij}+{\bar \beta}^{(2)}_{ij}
+{\cal O}(\alpha'^3)\ . \label{eq3.10}
\end{equation}
Combining (\ref{eq2.12}) and (\ref{eq3.8}), we obtain
\begin{eqnarray}
\frac{d{\cal S}}{ds} &=& \int_M \biggl \{ \frac{1}{4}{\bar
\beta}^{ij} \frac{\partial g_{ij}}{\partial
s}-\frac{\alpha'}{8}\bigg [ \nabla^i
{\bar \beta}^{jk}- \nabla^j {\bar \beta}^{ik}\bigg ]\nonumber\\
&& \qquad \left ( \nabla_i \frac{\partial g_{jk}}{\partial s} -
\nabla_j \frac{\partial g_{ik}}{\partial s} \right ) +\alpha'^3
T^{ij}\frac{\partial g_{ij}}{\partial s}\bigg \vert_{\varphi}\biggl
\} e^{-2\varphi}dV \ , \label{eq3.11}
\end{eqnarray}
where $T\in {\cal O}(1)$ is the coefficient of the error estimate.
Lastly, integrating by parts, we can express this in the form
\begin{equation}
\frac{d{\cal S}}{ds}=\int_M \frac{\partial g_{ij}}{\partial s}
g^{jl} \left \{ \frac{1}{4} \left [ g^{ik}\left ( 1+\alpha'
{\widetilde \Delta} \right ) -\alpha'{\widetilde {\rm
Hess}}^{ik}\right ]{\bar \beta}_{kl} +\alpha'^3T^i_l \right \}
e^{-2\varphi}dV\ . \label{eq3.12}
\end{equation}
Here ${\widetilde {\rm Hess}}^{ik}(\cdot):=e^{2\varphi}\nabla^k
\left ( e^{-2\varphi}\nabla^i (\cdot)\right )$ and $\widetilde
\Delta:=g_{ik}{\widetilde {\rm Hess}}^{ik}$.\footnote
{In other words, divergences are defined with respect to the measure
$e^{-2\varphi}dV$. At leading order in $\alpha'$, which is all that
we require here, this does not differ from the ordinary divergence
which appears in the comparable results in Section 2 of \cite{JJM}.}

\section{Gradient Flow and Monotonicity}
\setcounter{equation}{0}

\noindent In the finite-dimensional case, the flow
\begin{equation}
\frac{dx^i}{dt}=F^i \label{eq4.1}
\end{equation}
generated by vector field $F^i$ is a gradient flow iff for a metric
$\kappa$
\begin{equation}
F^i = \kappa^{ik}\partial_k V\ . \label{eq4.2}
\end{equation}
That is, $F$ is the gradient vector arising from raising the index
on the exterior derivative of a scalar potential $V$. Equivalently,
$F$ must obey
\begin{equation}
\partial_i F_j - \partial_j F_i =0 \quad , \quad F_i:=\kappa_{ik}F^k
\ . \label{eq4.3}
\end{equation}
The directional derivative of $V$ in the direction of an arbitrary
tangent vector $v^i=dx^i/ds$ is of course just
\begin{equation}
\frac{dV}{ds}=\frac{dx^k}{ds}\partial_k V\ . \label{eq4.4}
\end{equation}

In the infinite-dimensional case, the sums over $k$ become
integrals, the directional derivative (\ref{eq4.2}) becomes a
variational derivative, and $\partial_i V$ becomes measure-valued
(i.e., a distribution in the sense of Dirac). Given a candidate
potential function for a given flow generated by a known vector
field $F$, one can perform the variational derivative to read off
the analogue of $\partial_k V$ and then compare this to $F$ if the
metric is known. (When taking the variational derivative, the vector
field $dx^i/ds$ is replaced by the {\em co}tangent field $\partial
g_{ij}/\partial s$.)

We claim that a suitable metric $\kappa(\cdot,\cdot)$ on the space of couplings is
\begin{eqnarray}
\kappa(u,v)&:=&\frac{1}{4}\int_M dV\ e^{-2\varphi}\bigg [
g^{ik}g^{jl}u_{ij}v_{kl}-\frac{\alpha'}{2}g^{ik}g^{jl}g^{mn}\nonumber\\
&&\qquad \left ( \nabla_m u_{ij}-\nabla_i u_{mj} \right ) \left (
\nabla_n v_{kl}-\nabla_k
v_{nl}\right )\bigg ]\label{eq4.5}\\
&=&\int_M u_{ij}g^{jl} \biggl \{ \frac{1}{4} \left [ g^{ik} \left (
1+\alpha'{\widetilde \Delta} \right )-\alpha' {\widetilde {\rm
Hess}}^{ik}\right ] v_{kl}\nonumber\\
&&\qquad +{\cal O}(\alpha'^2|v|) \biggr \} e^{-2\varphi} dV \ ,
\label{eq4.6}
\end{eqnarray}
using integration by parts to obtain the last equality.

Compare (\ref{eq3.12}) to (\ref{eq4.6}). {\it Assuming} that
$T^{ij}$ is linear in ${\bar \beta}_{ij}$ and using
\begin{equation}
\frac{d{\cal S}}{ds}=-\kappa \left ( \frac{\partial g}{\partial s},
{\rm Grad\ }{\cal S} \right )\ , \label{eq4.7}
\end{equation}
then we can read off that the
gradient of ${\cal S}$ with respect to the metric $\kappa$ is
\begin{equation}
{\rm Grad\ } {\cal S}=-{\bar \beta}_{ij}+{\cal O}(\alpha'^3)\ ,
\label{eq4.8}
\end{equation}
establishing the claim.

For $u=v$, (\ref{eq4.5}) yields
\begin{equation}
\kappa(u,u)=\frac{1}{4}\int_M dV\ e^{-2\varphi}\left [ \vert u_{ij}
\vert^2-\frac{\alpha'}{2}\vert \nabla_i u_{jk}-\nabla_j u_{ik}
\vert^2 \right ] \ . \label{eq4.9}
\end{equation}
Notice the overall minus sign in front of the gradient terms.
Evaluating the derivative of ${\cal S}$ along a flow given
by (\ref{eq1.8}), the result (\ref{eq4.9}) leads to
\begin{eqnarray}
\frac{d{\cal S}}{dt}&=&-\kappa \left ( -{\bar \beta}, -{\bar\beta}
\right ) \nonumber\\
&=&-\frac{1}{4}\int_M dV\ e^{-2\varphi}\left [ \vert {\bar
\beta}_{ij} \vert^2-\frac{\alpha'}{2}\vert \nabla_i {\bar
\beta}_{jk}-\nabla_j {\bar \beta}_{ik} \vert^2 +{\cal
O}(\alpha'^4)\right ] \ . \label{eq4.10}
\end{eqnarray}

Thus as long as the nonderivative term dominates, ${\cal S}$ is
monotonically decreasing along the RG flow.

\section{Discussion}
\setcounter{equation}{0}

\noindent The action ${\cal S}$ will fail to be monotonically
decreasing at second order whenever
\begin{equation}
\vert \nabla_i {\bar \beta}_{jk}-\nabla_j {\bar \beta}_{ik} \vert >
\sqrt{\frac{2}{\alpha'}} \left \vert {\bar \beta}_{ij} \right \vert\
. \label{eq5.1}
\end{equation}
This situation is possible because we can always choose initial data
for the flow with spatial gradients that obey $\vert \nabla {\rm
Ric} \vert \sim \vert {\rm Ric} \vert /\sqrt{\alpha'}$. Then both
terms in the integrand of (\ref{eq4.10}) are comparable in
magnitude, and it may well be that the second term dominates, making
${\cal S}$ increase. However, when (\ref{eq5.1}) holds, then $\vert
\nabla {\rm Ric} \vert \sim \vert {\rm Ric} \vert /\sqrt{\alpha'}$.
Then the second order truncation of the $\beta$-function is no
longer reliable because third order terms are comparably large (cf
\cite{JJM}). The second order truncation breaks down. This scenario
and its possible outcomes were described in the Introduction.

We turn now to circumstances for which monotonicity does hold, at
least for an interval of ``time'' (energy scale).

One such class is the class of manifolds with harmonic curvature.
These are precisely the metrics for which
\begin{equation}
\nabla_k R^{ijkl}\equiv \nabla^i R^{jl} -\nabla^j R^{il}=0\ .
\label{eq5.2}
\end{equation}
Einstein manifolds obviously belong to this class, as do the
Riemannian products of Einstein manifolds (as these have parallel
Ricci tensor $\nabla_i R_{jk}=0$). Contracting this expression with
$g^{jl}$ and using the contracted second Bianchi identity, we see
that such manifolds must have constant scalar curvature.\footnote
{Quite a lot more is known about these metrics with harmonic
curvature; see \cite{Besse} and references therein.}
Then (\ref{eq2.5}--\ref{eq2.6}) admits solutions for $\Phi$ of the
form $1+{\cal O}(\alpha')$ and then $\varphi \in {\cal O}(\alpha')$
(i.e., $\varphi^{(0)}=0$). It follows that
\begin{equation}
\left [ \nabla_k {\bar \beta}^g_{ij}-\nabla_i {\bar
\beta}^g_{kj}\right ] \equiv \alpha' \left [ \nabla_k
R_{ij}-\nabla_i R_{kj} - R^l{}_{jki}\nabla_l \varphi\right ]
\in{\cal O}(\alpha'^2)\ , \label{eq5.3}
\end{equation}
assuming ${\cal O}(1)$ bounds on the curvature. Provided the
solution is not ``nearly solitonic'' (i.e., provided ${\bar
\beta}^{(1)}\notin {\cal O}(\alpha'^2)$), then $d{\cal S}/dt<0$.

The condition of harmonic curvature cannot be expected to be
preserved along the flow in general. If $t$ becomes large enough, an
initially harmonic curvature can eventually deviate quite a bit from
harmonicity.

A second class that obeys monotonicity is the class of gradient
Ricci solitons, including so-called shrinkers and expanders as well
as steadies.\footnote{
Since we work with compact manifolds, the steady solitons are all
Ricci-flat \cite{Bourguignon}.}
These obey
\begin{equation}
{\bar \beta}^{(1)}_{ij} \equiv \alpha' \left ( R_{ij} + 2 \nabla_i
\nabla_j \varphi \right ) = \alpha' \lambda g_{ij} \ , \label{eq5.4}
\end{equation}
where $\lambda$ is a constant. Clearly, for this class, $\nabla_i
{\bar \beta}^{(1)}_{jk}=0$, so the wrong-sign term in (\ref{eq4.10})
vanishes, while the leading term integrates to give
\begin{equation}
\frac{d{\cal S}}{dt} = -\frac{n\lambda^2\alpha'^2}{4} + {\cal O}
(\alpha'^3)\label{eq5.5}
\end{equation}
in dimension $n$, where ${\cal O} (\alpha'^3)$ denotes the
contribution from ${\bar \beta}^{(2)}_{ij}$. Again, this class will
not be preserved along the flow, but deviations will be governed by
the $\alpha'^2 R_{iklm}R_j{}^{klm}$ term in (\ref{eq1.8}), and such
deviations, if absent initially, will not be important for quite
some time. In fact, all that is required is that the evolving metric
have gradient of ${\bar \beta}_{ij}$ close to that of a soliton
metric; i.e., close to zero.

\section{Acknowledgments}

\noindent We are grateful to Arkady Tseytlin for his comments on a
draft of this paper. EW would like to thank the Dept of Mathematics
and Statistics, University of New Brunswick, for hospitality during
the beginning of this work; the Albert Einstein Institute for an
invitation to the Workshop on Geometric and Renormalization Group
Flows, whose stimulating atmosphere led to the work's completion;
and H-P K\"unzle for a discussion of manifolds of harmonic
curvature. This work was partially supported by a grant from the
Natural Sciences and Engineering Research Council of Canada.

\end{document}